\documentclass[aps,floatfix,prl,twocolumn,a4paper,10pt,citeautoscript,
preprintnumbers,longbibliography,showpacs,nofootinbib]{revtex4-1}

\usepackage[english]{babel} 
\usepackage{amsmath}        
\usepackage{amssymb}        
\usepackage{bm}         
\usepackage{bbm}            

\usepackage{graphicx}       
\DeclareGraphicsExtensions{.pdf}    

\usepackage[colorlinks=true, pdfstartview=FitV,linkcolor=blue,
            citecolor=blue, urlcolor=blue]{hyperref}
\newcommand{\Align}[2]{\begin{align}\label{#1}#2\end{align}}

\newcommand{\bs}{\boldsymbol}

\newcommand{\Figref}[1]{Fig.~\ref{#1}}
\newcommand{\Eqref}[1]{\eqref{#1}}

\newcommand{\groupU}[1]{{U}(#1)}   
\newcommand{\groupZ}[1]{\mathbb{Z}_{#1}} 

\newcommand{\Grad}{{\bs\nabla}}

\newcommand{\Curl}{{\bs\nabla}\times}

\newcommand{\A}{{\bs A}}
\newcommand{\B}{{\bs B}}
\newcommand{\J}{{\bs J}}

\newcommand{\F}{\mathcal{F}}

\newcommand{\Tz}{T_{\groupZ{2}}  }

\begin{document}
\title{Unconventional thermoelectric effect in superconductors
that break time-reversal symmetry}

\author{Mihail~Silaev}
\author{Julien~Garaud}
\author{Egor~Babaev}
\affiliation{Department of Theoretical Physics, KTH-Royal
Institute of Technology, Stockholm, SE-10691 Sweden}

\begin{abstract}
We demonstrate that superconductors which break time-reversal
symmetry can exhibit thermoelectric properties, which are entirely
different from the Ginzburg mechanism. As an example, we show
that in the $s+is$ superconducting state there is a reversible
contribution to thermally induced supercurrent, whose direction
is not invariant under time-reversal operation. Moreover in
contrast to Ginzburg's mechanism it has a singular behavior near
the time-reversal symmetry breaking phase transition. The effect
can be used to confirm or rule out the $s+is$ state, which is
widely expected to be realized in pnictide compounds
Ba$_{1-x}$K$_{x}$Fe$_2$As$_2$ and stoichiometric LiFeAs.
\end{abstract}
\date{\today}
\maketitle

Thermoelectric effects in superconductors, and their fundamental
importance, were first discussed 70 years ago by Ginzburg
\cite{GinzburgOld1,GinzburgOld2}. Currently there is a 
revival of interest in this topic \cite{FSThermophase1,
FSThermophase2,FSThermophase3,Loefwander.Fogelstroem:04}.
These effects originate in charge transfer by thermal
quasiparticles \cite{VanHarlingen}, which is compensated by the
counterflow of superconducting current $\bm{j}_s=-b_n\Grad T$
(where $b_n$ is the normal component thermoelectric coefficient).
As it is determined by the dissipative normal current, such a
thermally induced supercurrent is irreversible since $\bm{j}_s$
changes sign under the time-reversal transformation while $b_n$
and $\Grad T$ remain invariant.
We point out here, that multicomponent superconductors which break
the time-reversal symmetry, have entirely different reversible
thermoelectric response.  Namely, there is a \emph{generic}
contribution to the thermally induced supercurrent, whose
direction is not preserved under time-reversal operation. Besides,
that contribution can be orders of magnitude larger than the
ordinary one and, in contrast to Ginzburg mechanism, be present
even at low temperatures. More precisely the new contribution we
describe is important in the vicinity of the time-reversal symmetry
breaking phase transition, that can occur at temperatures much
lower than $T_c$, where the usual contribution due to thermal
quasiparticles is typically extinct.
To illustrate this new effect we consider time-reversal symmetry 
breaking states which are widely expected to exist in iron pnictide 
compounds  \cite{Chubukov1,ChubukovLiFeAs,Lee.Zhang.Wu:09,Zhang2}.

\begin{figure*}[htbf]
\hbox to \linewidth{ \hss
\includegraphics[width=.9\linewidth]{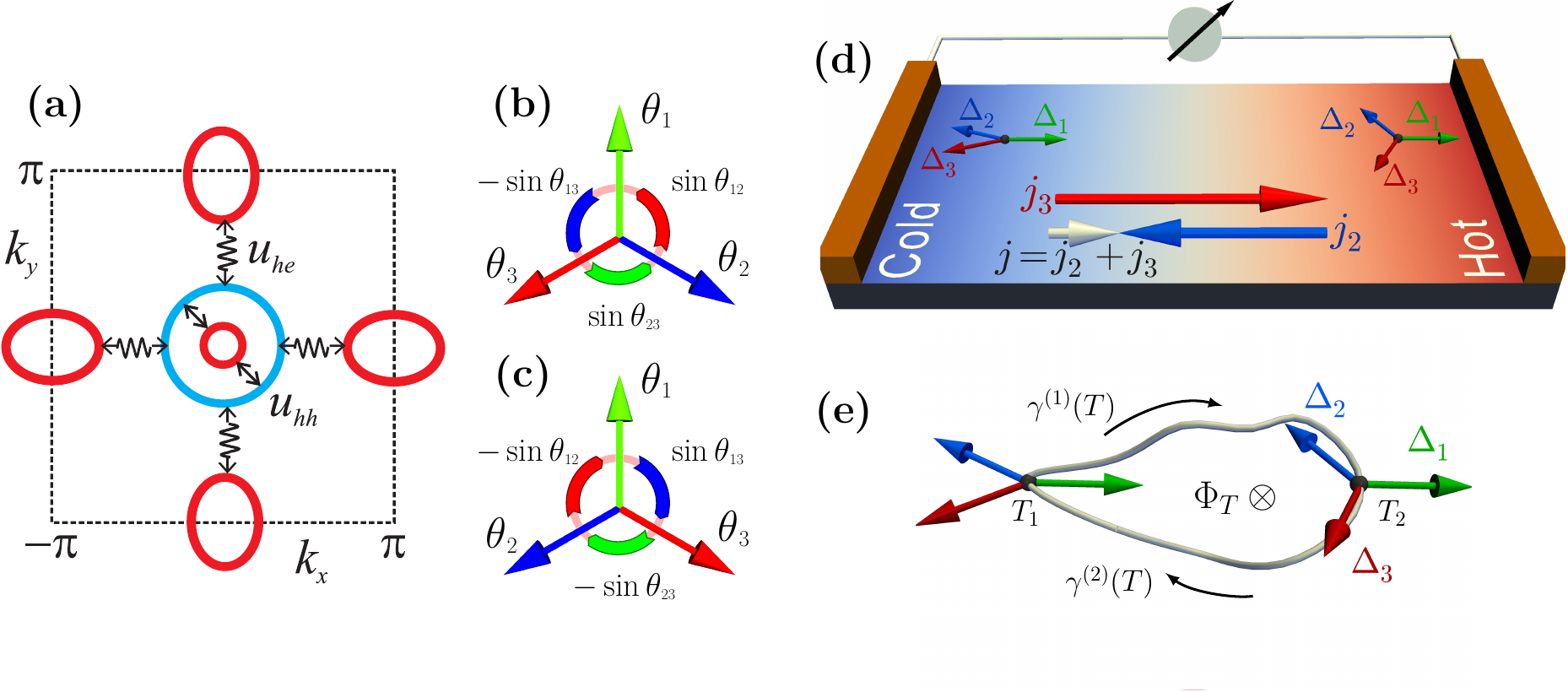}
\hss}
\vspace{-1cm}
\caption{ (Color online) --
(a)  Schematic view of the band structure in hole-doped iron
pnictide compound Ba$_{1-x}$K$_{x}$Fe$_2$As$_2$. It consists of
two hole pockets at $\Gamma$ shown by open circles and two
electron pockets at $(0,\pi)$ and $(\pi, 0)$ shown by ellipses.
The s+is state is favored by superconducting coupling dominated by
the interband repulsion between electron and hole Fermi surfaces,
as well as between the two hole pockets.
Panels (b,c) schematically show the two degenerate ground-states
$s\pm is$. The phases of the order parameter components
$\Delta_k=|\Delta_k|e^{i\theta_k} $ are represented by vector
diagrams. Circular arrows show the directions of interband
Josephson current supported by non-trivial relative phases
$\theta_{kj}\neq 0$, $\pi$.
(d) Sample whose ends are maintained at different temperatures $T_{1,2}$.
The thermophase effect appears due to the temperature-dependent
interband phase difference $\theta_{kj}=\theta_{kj}(T)$. The total
thermally induced current ${\bm j}$ will have opposite directions
in $s\pm is$ states.
Panel (e) sketches a closed circuit in spatially inhomogeneous bulk
superconductor with two branches characterized by different values
of the thermophase coefficients $\gamma^{(1,2)}(T)$. The junctions
between branches have different temperatures $T_{1,2}$ and the
temperature bias generates magnetic flux through the circuit
\Eqref{Eq:ThermalFlux1}.
}
\label{Fig:Fig0}
\end{figure*}

A fundamentally interesting state that can appear in multicomponent
superconducting systems is the so called $s+is$ state. In addition
to the usual gauge symmetry $U(1)$, it is characterized by a broken
time-reversal symmetry (BTRS).
Not only is it the simplest BTRS extension of the most abundant $s$-wave
state, but also it is expected to arise from various microscopic physics 
\cite{StanevTesanovic,Thomale,Suzuki,Chubukov1,Chubukov2,ChubukovLiFeAs}.
The $s+is$ state can as well be fabricated on demand on the interfaces
of superconducting bilayers \cite{Ng}. Recently it was demonstrated
that such physics very likely occurs in strongly hole doped
Ba$_{1-x}$K$_{x}$Fe$_2$As$_2$ \cite{Chubukov1,Chubukov2} as well as
in stoichiometric compound LiFeAs \cite{ChubukovLiFeAs}.

In iron pnictides, the $s+is$ BTRS state originates from the
multiband character of superconductivity and several competing
pairing channels, as shown schematically on the inset in
\Figref{Fig:Fig0}(a). A typical band structure of
Ba$_{1-x}$K$_{x}$Fe$_2$As$_2$ consists of two hole pockets at the
$\Gamma$ point and two electron pockets at $(0,\pi)$ and $(\pi,
0)$. The superconducting coupling here is dominated by the
interband repulsion between electron and hole Fermi surfaces, as
well as between the two hole pockets at $\Gamma$. Such a system
can be described by a minimal three-band model, assuming that the
order parameter components are $\Delta_1$ for electron pockets and
$\Delta_{2,3}$ for the hole bands at $\Gamma$ \cite{Chubukov1}.
 The competition of repulsion forces results in
an intrinsically complex order parameter whose components in each
band $\Delta_k=|\Delta_k|e^{i\theta_k}$ (where $k=1,2,3$ is the
band index) possess non-trivial (frustrated) ground-state phase
differences $\theta_{kj}=\theta_k-\theta_j $ which are neither $0$
nor $\pi$ \cite{StanevTesanovic,Thomale,Suzuki,Chubukov1,Chubukov2,Johan}.
The ground-state order parameter of the $s+is$ phase is not invariant
under the discrete ($\groupZ{2}$) time-reversal transformation associated
with complex conjugation ${\cal T}(\Delta_k)= \Delta_k^*$. Thus it has
a broken $\groupU{1}\times\groupZ{2}$ symmetry \cite{Johan} and below,
$s\pm is$ will stand for states that are related by time-reversal
transformation ${\cal T}$. One of the interband phase differences,
e.g. $\theta_{12}$ can be used as the order parameter that
characterizes the $\groupZ{2}$ phase transition.

The BTRS states with phase frustration $\theta_{kj}\neq0$, $\pi$
imply the existence of persistent ``intrinsic Josephson-like
currents" between the three bands, as shown schematically in
\Figref{Fig:Fig0}(b,c). These currents have opposite ``directions" 
for the two ground states with opposite values of interband phase 
differences. Although the underlying microscopic physics is not exotic, 
nor is any fine tuning is required to form such a state, no experimental
observations of such BTRS states have been reported. The reason
being the major challenges to distinguish the $s+is$ state from
its time-reversal invariant $s_\pm$ and $s_{++}$ cousins by conventional 
methods. Indeed, the probe of relative phases between components of the 
order parameter in different bands is a non-trivial task. Few proposals
have recently emerged for the indirect observation of BTRS signatures in 
collective mode spectrum \cite{BlumbergMgB2,Lin,Stanev,Benfatto,Chubukov2,
ChubukovRaman,Johan}, unusual topological defects \cite{Garaud.Carlstrom.ea:11,
Garaud.Babaev:14} or spontaneous currents near impurities in samples 
subjected to strain \cite{ChubukovMaitiSigrist}.

The challenges associated with the observation of the $s+is$ state
and its relevance to pnictides motivate our choice of this kind of
material for demonstration of the unusual thermoelectric
properties of BTRS superconductors. Below we show that in such
systems, there is a new mechanism for thermal generation of
supercurrent. It is not related to normal thermal currents and
leads to thermoelectric properties which strongly differ from that
suggested by Ginzburg, in usual superconductors. Although the suggested
mechanism does not eliminate the usual thermoelectric effect, it will
be shown to dominate near the BTRS phase transition thus providing
an experimental test for this unconventional state.

The key idea behind the proposed thermoelectric effect is that,
due to the generically temperature-dependent interband phase
differences, i.e., $\theta_{kj}=\theta_{kj}(T)$, a temperature
bias generates phase gradients of condensate components.
Assuming that temperature gradients are small, so that the
order parameter is determined by the local temperature, yields
the relation
\begin{equation}\label{Eq:Phase}
  \Grad\theta_{k} = \Grad\varphi_0+\gamma_k (T)\Grad T\,,
\end{equation}
where $\varphi_0=\sum_{k=1}^3\theta_k/3$,  and $\gamma_k (T)=
\frac{1}{3}\sum_{j\neq k}d\theta_{kj}/dT$ are the
\emph{thermophase coefficients}.
Superconducting currents can be generated, as a results of this
\emph{thermophase effect}. In multiband superconductors, the total
current has contributions from each band ${\bm j} = \sum_k{\bm j}_k$.
In units where $\hbar=c=1$, the partial currents read as
 ${\bm j}_{k}=(\Grad\theta_{k}-2\pi\A/\Phi_0)/8e\pi\lambda_{k}^2$,
where $\lambda_{k}$ are characteristic constants associated with
contributions of a given band to the London penetration depth.
$\A$ is the vector potential of the magnetic field $\B=\Curl\A$
and $\Phi_0$ the flux quantum.
Introducing the notation ${\bm Q}_0= \Grad\varphi_0 -2\pi\A/\Phi_0$
and the London penetration length $\lambda^2_L=1/\sum_k \lambda^{-2}_{k}$,
allows to write the total current as
\begin{equation}\label{Eq:GLCurrent}
 {\bm j} = \frac{{\bm Q}_0}{8e\pi\lambda_L^2} +
 \sum_{k>i} \frac{\lambda^{2}_{i}  -
 \lambda^{2}_{k}}{24e\pi \lambda^{2}_{k}\lambda^{2}_{i}} \Grad \theta_{ki}.
\end{equation}
The first term here is a usual Meissner current while the
second part is determined by the gradients of interband phase
differences. It describes the charge transfer by counter-currents
of several superconducting components.
Observe that  according to Eq.~\Eqref{Eq:Phase}, the counterflow
is generated by the temperature gradient and contributes to the
current according to the following generic expression:
\begin{align}\label{Eq:TotalCurrent}
 {\bm j} &= \frac{ {\bm Q}_0 + \gamma(T) \Grad T}{8e\pi\lambda_L^2(T)}
 \\ \label{Eq:gamma}
 \gamma(T) &= \lambda_L^2(T)\sum_k \gamma_k (T) \lambda_{k}^{-2}(T) .
\end{align}
The essence of such a thermophase effect is schematically shown in
\Figref{Fig:Fig0}(d), for the case of  ${\bm Q}_0=0$ so that,
according to Eq.~\Eqref{Eq:Phase}, the partial currents in each
band are given by ${\bm j}_{k}= \gamma_k (T) \Grad T /
8e\pi\lambda_k^2$. Since the thermophase coefficients are opposite
for the $s\pm is$ states: ${\cal T} \gamma_k=-\gamma_k $, these
thermally induced superconducting currents are sensitive to
time-reversal symmetry transformation. That is, for a given
temperature bias, the current directions are opposite in $s+is$
and in $s-is$ states.

Deep in the bulk of a superconducting sample the total current
\Eqref{Eq:TotalCurrent} should vanish: ${\bm j}=0$, which yields
the condition for current compensation
\begin{equation}\label{Eq:Compensation}
{\bm Q}_0=-\gamma\Grad T \,.
\end{equation}
 This compensation can be achieved locally even for
inhomogeneous superconductor if the thermophase coefficient has
slow spatial dependence $\gamma=\gamma(T,{\bm r})$. Integrating
Eq.~\Eqref{Eq:Compensation} along a closed path and assuming for
simplicity that no vortices are trapped in the circuit ($\oint
{\bs{d\ell}}\cdot\Grad\varphi_0 =0$), we get the thermally induced
magnetic flux
\begin{equation}\label{Eq:ThermalFlux}
\Phi_T= \frac{\Phi_0}{2\pi} \oint {\bs{d\ell}}\cdot\Grad T
\gamma(T,{\bm r}) \,.
\end{equation}
Since the coefficient $\gamma(T,{\bm r})$ is in general spatially
inhomogeneous along the integration path, the expression
\Eqref{Eq:ThermalFlux} yields a finite value of $\Phi_T$.

Now consider a closed circuit with two branches that are made up of
different superconductors, for example the
Ba$_{1-x}$K$_{x}$Fe$_2$As$_2$ compound with different doping
levels. The thermophase coefficient $\gamma(T)$ in this case has
a step-wise discontinuity along the circuit, determined by the
values $\gamma^{(1,2)}(T)$ at the different branches. This is
shown schematically in \Figref{Fig:Fig0}(e). Assuming that the
junctions between branches have different temperatures $T_{1,2}$,
Eq.~\Eqref{Eq:ThermalFlux} yields an induced magnetic flux through
the circuit given by
\begin{equation}\label{Eq:ThermalFlux1}
\Phi_T= \frac{\Phi_0}{2\pi} \int_{T_1}^{T_2} d T [ \gamma^{(1)}(T)
- \gamma^{(2)} (T) ]   \,.
\end{equation}
The maximal possible magnitude of $\Phi_T$ can be estimated by
considering one of the branches to be in the time-reversal
invariant state, so that e.g. $\gamma^{(2)}(T)=0$ and
$\lambda_1\ll\lambda_2$. Then, from Eq.~\Eqref{Eq:ThermalFlux1},
we obtain $\Phi_T/\Phi_0 \approx\delta\theta_{12}/4\pi $ where
$\delta\theta_{12}=\theta_{12}(T_2)- \theta_{12}(T_1) $.

The resulting thermally induced flux can have a giant magnitude
compared to that produced by a usual thermoelectric effect. Below,
we will  introduce a realistic microscopic model for the $s+is$
state, to demonstrate that the interband phase difference can have
a significant variation as a function of temperature
$\max(\delta\theta_{12})\sim\pi$. Therefore the resulting flux
$\Phi_T \sim \Phi_0$ is much larger than the typical value of
$\Phi_T \sim  10^{-2}\Phi_0$ expected in usual superconducting
thermoelectric circuits \cite{GinzburgOld1,GinzburgOld2,Kozub1,
Kozub2,Zavaritskii:74}
\footnote{ The magnitude of quasiparticle thermoelectric
effects in superconductors is still a challenging problem,
see discussion in Ref.~\cite{Galperin.Gurevich.ea:02}. It is beyond
the scope of the present paper since we deal with unconventional
thermoelectric effect, which is not related to the flow of
quasiparticles.
}. \nocite{Galperin.Gurevich.ea:02}

To provide a microscopic basis for this physics, we now proceed
to calculate thermophase coefficients for the $s+is$
superconducting state in Ba$_{1-x}$K$_{x}$Fe$_2$As$_2$. Within
the minimal three-band model, it is parametrized by two pairing
constants characterizing the strength of interband hole-hole
$u_{hh}$ and electron-hole $u_{he}$ repulsions  (as shown
schematically in \Figref{Fig:Fig0}).
%
 The self-consistency equation for the order parameter
components ${\bm \Delta} = ( \Delta_{1}, \Delta_{2}, \Delta_3)$
\cite{Benfatto,Chubukov1,Chubukov2}
\begin{equation}\label{Eq:SelfConsistency1}
{\bm \Delta } =2\pi T\hat\Lambda  \sum_{n=0}^{N_d} {\bm F
(\omega_n) }\,,
\end{equation}
where  ${\bm F} = ( F_{1}, F_{2}, F_{3} )$  and $F_{k} (\omega_n)=
\Delta_k/\sqrt{\omega_n^2+ |\Delta_k|^2} $, $\omega_n=(2n+1)\pi T$
are fermionic Matsubara frequencies, $T$ is the temperature. Here
$N_d=\Omega_d/(2\pi T)$ is a cut-off at Debye frequency, and the
coupling parameter matrix is chosen in the form
 \begin{equation}\label{Eq:Model3Band}
\hat\Lambda = - \left(%
\begin{array}{ccc}
  0 & u_{hh} & u_{he} \\
  u_{hh} & 0 & u_{he} \\
  u_{he} & u_{he} & 0 \\
\end{array}%
\right) \,,
\end{equation}
which yields the phase frustration for certain parameters
$u_{hh}$, $u_{he}$. In this case $\Delta_{1,2}$ and $\Delta_3$
describe the order parameter in hole bands and electronic pockets
correspondingly. The order parameter has symmetric form ${\bf
\Delta} = e^{i\varphi_0} ( |\Delta_1| e^{i\theta_{12}} ,
|\Delta_2| e^{-i\theta_{12}}, |\Delta_3| )$, where $\varphi_0$ is
the common phase and the non-trivial interband phase difference
$\theta_{12} \neq 0, \pi$ appears in the TRSB $s+is$ state.
%
\begin{figure}[!hb]
\hbox to \linewidth{ \hss
\includegraphics[width=\linewidth]{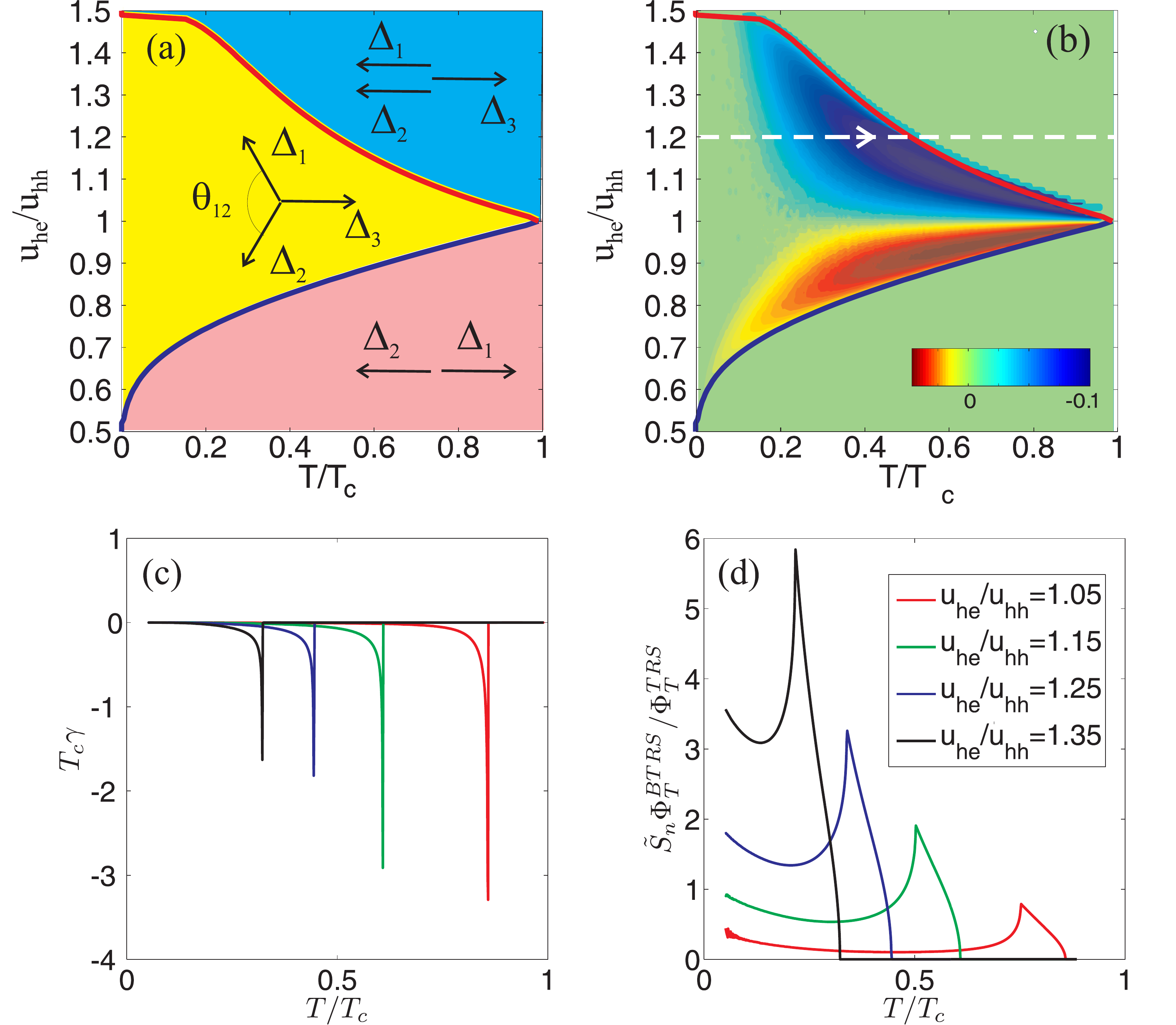}
\hss} \caption{ (Color online) -- (a) Phase diagram of the
three-band superconductor described by the microscopic model
(see text). Solid curves show the critical temperature $\Tz$
of the BTRS phase transition.
(b) Color scale plot of the thermophase coefficient $(\Tz-T)\gamma(T)$.
The prefactor $(\Tz-T)$ is added to regularize the square-root
singularity of $\gamma(T)$ at the BTRS phase transition.
(c) The temperature dependencies of $\gamma(T)$ along the lines
such as shown by arrow in the panel (b). From right to left
$u_{he}/u_{hh}=1.05;\; 1.15;\; 1.25;\; 1.35$. (d) The ratio of
magnetic fluxes generated due to the unconventional and
conventional thermoelectric effects, $\Phi_T^{BTRS}$ and
$\Phi_T^{TRS}$ correspondingly. $\tilde{S}_n$ is a dimensionless
normal state Seebeck coefficient which can be of the order $1$ but
typically is much smaller. The other system parameters are
explained in text.
}
\label{Fig:PhaseDiagram}
\end{figure}
%
The phase diagram, shown in \Figref{Fig:PhaseDiagram}, is obtained
by solving a nonlinear self-consistency problem
(\ref{Eq:SelfConsistency1},\ref{Eq:Model3Band}). It demonstrates
the line of $\groupZ{2}$ BTRS phase transition that occurs at the
temperature $\Tz\leq T_c$. Within this model the thermophase
coefficients in Eq.~\Eqref{Eq:TotalCurrent} are given by $\gamma_1
(T) = - \gamma_2 (T) = - \frac{1}{2}d\theta_{12}/dT$ and $\gamma_3
=0$. Despite the symmetry between bands the net thermophase
coefficient \Eqref{Eq:gamma} is in general non-zero since the
coefficients $\lambda_k$ depend not only on the gap amplitudes but
also on the kinetic coefficients. For example in the diffusive
limit they are given by a standard text-book expression for the
London length $\lambda^{-2}_k= 4\pi^2 \sigma_k
|\Delta_k|\tanh(|\Delta_k|/2T)$ where $\sigma_{k}$ are the normal
state conductivities in each band which are different even if the
superconducting pairing is symmetric between hole bands.
In the vicinity of $\Tz$ the thermophase coefficients diverge as
$\gamma(T) \sim(\Tz-T)^{-1/2}$.  In \Figref{Fig:PhaseDiagram}(b)
the overall profile of the function $(\Tz-T)\gamma(T)$ is shown,
where the prefactor $(\Tz-T)$ is added to remove the divergence.
The typical temperature dependencies of $\gamma(T)$ are shown in
\Figref{Fig:PhaseDiagram}(c) for particular coupling parameters.
The square-root singularity dependence of thermophase coefficient
(see \Figref{Fig:PhaseDiagram}c) is consistent with the fact that
the effective potential for the phase difference gets soft near
the $\groupZ{2}$ symmetry-restoring phase transition. However it
is in strong contrast with the quasiparticle contribution
which has a smooth thermally activated
behavior\cite{Kozub1,Kozub2}. 

Hence the thermophase coefficients are large near $\Tz$ and the
discussed thermoelectric effect is generically the dominating one
at low temperatures in the fully gapped BTRS $s+is$ state. The
magnitude of the total variation of the relative phase $\theta_{12}$
as a function of temperature can be estimated from the typical
dependence in \Figref{Fig:PhaseDiagram}(c). There
$\delta\theta_{12}\approx 0.4\pi$, so that the thermally generated
flux is $\Phi_T \sim 0.2\Phi_0$ which is in general much
larger than the theoretically predicted value for quasiparticle
contribution.

To draw a more convincing comparison let us calculate the
thermoelectric fluxes in the BTRS state $\Phi^{BTRS}_T$ and in the 
usual time-reversal symmetric superconductor $\Phi^{TRS}_T$ generated 
in the thermoelectric circuit according to
Eq.\Eqref{Eq:ThermalFlux1}. For simplicity we assume that one of
the branches is passive $\gamma_2=0$ and considerer the
temperature bias $T_2-T_1 =0.1 T_c$ which is of the order of $1-5$
K in iron pnictides. To calculate a usual quasiparticle
contribution we use the result of Refs.\cite{Kozub1,Kozub2}:
$\gamma = 4\pi\lambda_L^2 \eta$ where $\eta = 0.6
\sigma_n\tilde{S}_n (T/T_c) \int^{\infty}_{\Delta_0/T}dy y^2
\cosh^{-2}(y/2)$. Here $\Delta_0$ is the smallest gap in the
system, $\sigma_n$ is the normal state conductivity and
$\tilde{S}_n = (e/k_B) S_n$ where $S_n$ is a Seebeck coefficient
just above $T_c$, $k_B$ is Boltzmann constant and $e$ is the
electron charge. We use  a diffusive model for $\lambda_k$
parametrized by kinetic coefficients
$\sigma_3=\sigma_2=\alpha\sigma_1$ where the coefficient
$\alpha\sim 1$ measures an asymmetry between hole pockets. Note
that within our simplified model of equal coupling constants for
the hole bands the overall magnitude of the unconventional
thermoelectric effect is proportional to the asymmetry of kinetic
coefficients $ \sigma_1-\sigma_2 = (\alpha-1)\sigma_1$, which we
assume below to be rather small $\alpha =0.9$. For such parameters
the typical temperature dependencies of the ratio
$\tilde{S}_n\Phi_{BTRS}/\Phi_{TRS}$ are shown in
\Figref{Fig:PhaseDiagram}(d). From this plot one can see that even
for quite a large values of Seebeck coefficient $S_n \sim 100 \mu
V/K$ and hence $\tilde{S}_n \sim 1$ observed in some
iron-oxipnictide compounds \cite{Exp:ThElPnictide} we obtain that
the unconventional thermoelectric signal is in fact much larger
than the quasiparticle contribution especially if $T_{Z2}\ll T_c$,
see black curve in \Figref{Fig:PhaseDiagram}(d).

The usual thermoelectric effect can be significant at low
temperatures if quasiparticle density remains finite, e.g., in
gapless superconductors. However, we would like to emphasize that
even in this case the suggested effect can be separated from
quasiparticle contribution since it breaks the time-reversal
symmetry. That is, the direction of thermally induced magnetic
field is different in $s+is$ and $s-is$ states while the usual
thermoelectric effect is time-reversal invariant.

Let us now  consider several characteristic examples which should
arise in concrete experimental set-ups based on circuits
containing BTRS $s+is$ superconductors. To that end, we use the
three-component Ginzburg-Landau (GL) theory \Eqref{Eq:GL},
although the effects will be similar in the microscopic formalism.
The GL expansion, derived from the microscopic three-band model
(see Appendix~\ref{Appendix1} for details), reads as
 \begin{align}\label{Eq:GL}
 {\cal F}&= \sum_{k=1}^3 \frac{D_k}{2} 
\Big|\Big(\Grad+i\frac{2\pi}{\Phi_0}\A\Big)\Delta_k\Big|^2
+\alpha_k|\Delta_k|^2+\frac{\beta_k}{2}|\Delta_k|^4 \nonumber \\
&+\sum_{k=1}^3\sum_{j>k}^3\eta_{kj}|\Delta_k||\Delta_j|\cos\theta_{kj}
+\frac{\bm B^2}{2} \,,
 \end{align}
where $j,k$ are band indices, and $D_k$ are diffusion constants.
We model temperature dependence of the coefficients as
$\alpha_k\simeq\alpha^0_k\left( T/T_k -1\right)$ ($\alpha^0_k$
and $T_k$ being characteristic constants, further details on the
derivation and choice of parameters are given in Appendix~\ref{Appendix1}).
%
\begin{figure}[!htb]
\hbox to \linewidth{ \hss
\includegraphics[width=\linewidth]{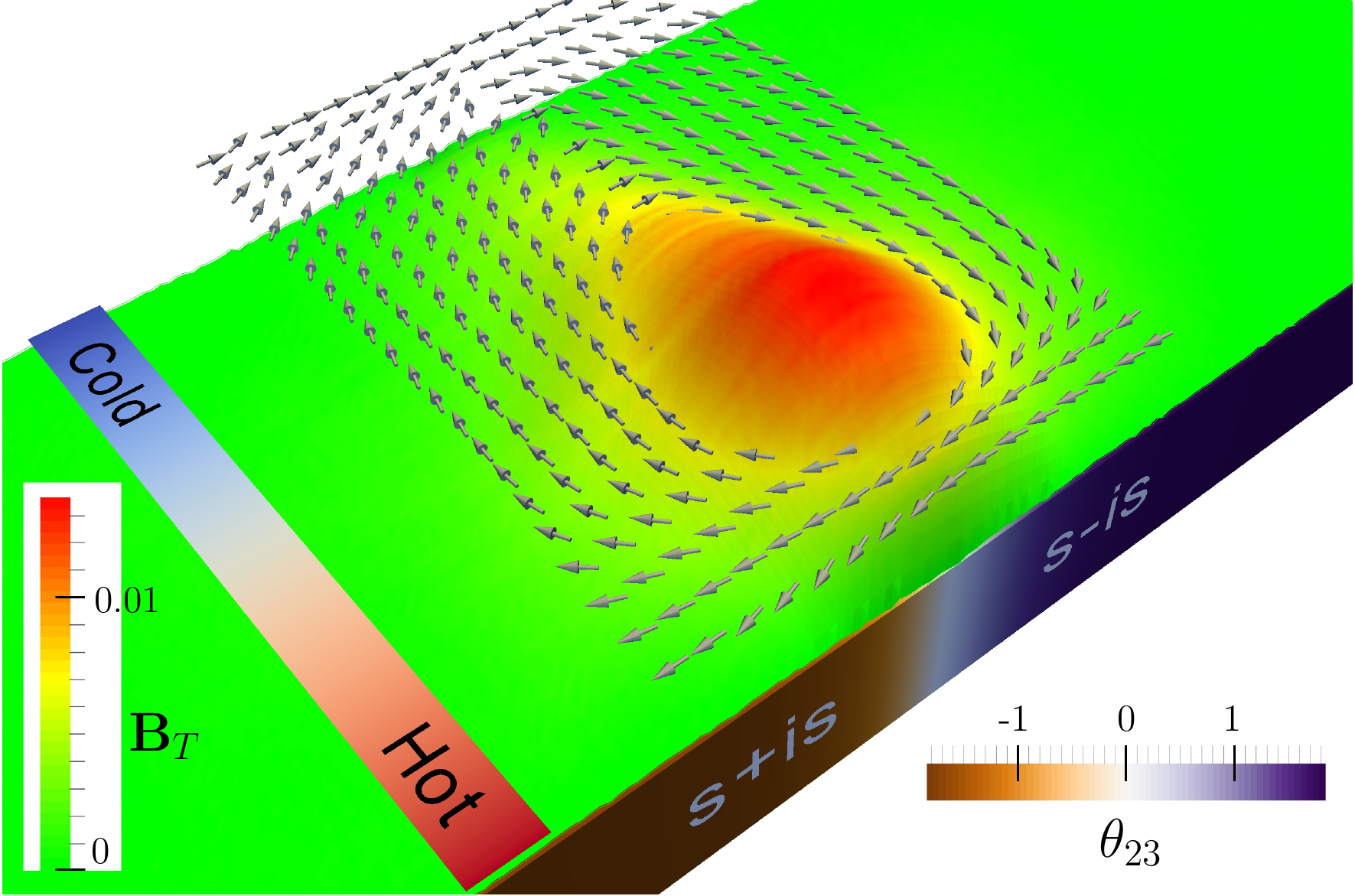}
\hss}
\caption{ (Color online) --
Thermally induced magnetic field on a domain-wall between $s+is$
and $s-is$ phases. Two faces of the stripe are maintained at
different temperatures, thus resulting in a thermal gradient ,
here $\delta T=0.2T_c$, along the domain-wall.
The $s+is/s-is$ domain-wall generates the magnetic field
${\bf B}_T$. In the absence of the temperature gradient a domain-wall
in $s+is$ superconductor does not have any magnetic signature.
}\label{Fig:DW}
\end{figure}
%
The numerical studies of the GL equations are performed within
a finite-element formulation \cite{Hecht:12} and using a non-linear
conjugate gradient algorithm, with the standard condition that no
current flows through the sample's boundaries.

\vspace{0.5cm}
First, we consider thermoelectric response of a sample with a domain 
wall that separate $s\pm is$ phases,  which are natural inhomogeneities
expected to exist in realistic systems with spontaneously broken
discrete symmetries.
It is known that such domain walls do not carry magnetic field at
constant temperature \cite{Garaud.Babaev:14}. We find that, in the
presence of a temperature gradient along the domain wall, it
generates a thermally induced magnetic field, as shown in
\Figref{Fig:DW}. There, the thermophase coefficients $\gamma(T)$
have opposite signs in $s+is/s-is$ domains. Therefore, in the
vicinity of the interface between these, there should be net
superconducting current and a thermally induced magnetic field
${\bm B}_T$.
Note that there is no phase winding in this flux-carrying configuration
and the thermally induced flux ${\bm B}_T$ is not quantized
\footnote{
Magnetic flux penetration into a superconductor without phase
winding/vortices is not restricted to three-band superconductors.
This can be seen from the general relation for magnetic flux
obtained from the expression supercurrent $\Phi=\oint \A d\ell =
\oint [\J/\rho({\bf r}) + \sum_{i=1..N} |\Delta_i ({\bf
r})|^2\nabla\theta_i( {\bf r})/\rho({\bf r})] $ ($\rho({\bf r})$
being the total density). Even if there is no phase winding and
even if we select the integration contour where $\J \to 0$, the
integral will in general be nonzero, in the presence of gradients
of relative phases $\theta_i-\theta_j$ and densities.
}.
Since the condensate density near domain wall is
non-homogeneous one cannot avoid additional signal produced by
usual the thermoelectric effect according to Ginzburg's mechanism.
However, it generates a dipole-like magnetic field distribution
which does not affect the total thermally induced magnetic flux.
For a homogeneous BTRS state in the absence of domain walls both
normal and unconventional thermoelectric signals are absent.

\begin{figure}[!htb]
\hbox to \linewidth{ \hss
\includegraphics[width=0.9\linewidth]{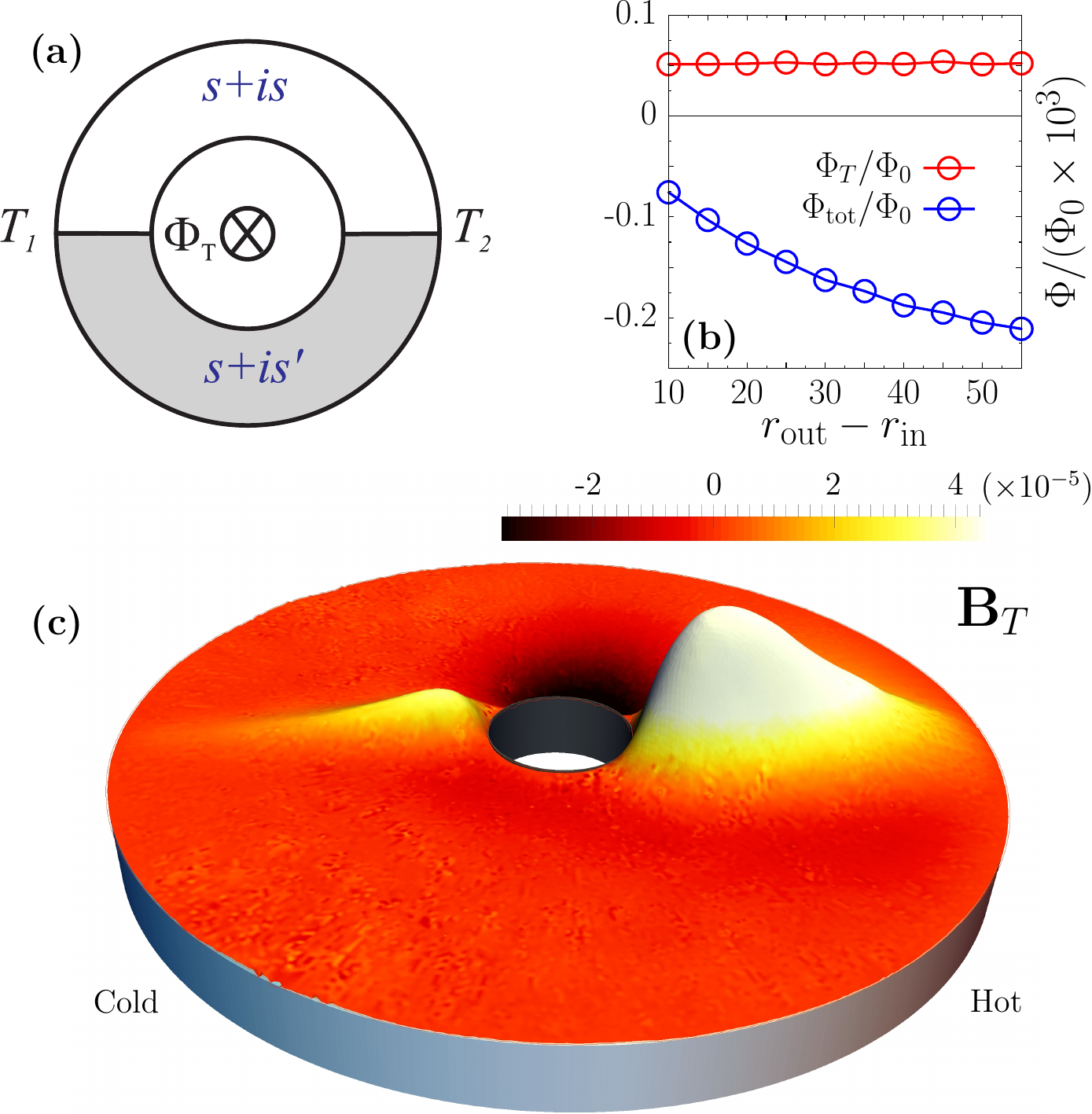}
\hss} 
\caption{ (Color online) -- A bimetallic
ring consisting of two branches in the $s\pm is$ state, but having
different chemical composition (see details in
Appendix~\ref{Appendix2}). Both junctions are maintained at
different temperatures $T_{1,2}$, so that there is a thermal
gradient that varies linearly with the polar angle. This is
sketched in panel (a).
The resulting thermally induced magnetic field is shown on panel
(c). In contrast to the usual thermoelectric effect its overall
sign depends not only on the temperature bias but also on which of
$\groupZ{2}$ BTRS ground states (i.e. $s+is$ or $s-is$) is
realized.
This configuration supports nonzero flux $\Phi_T$ in the inner
hole, which is more or less independent of the diameter of the
inner hole. The total flux, on the other hand depends on the
diameter of the sample (b).
}\label{Fig:Ring}
\end{figure}

The non-trivial thermoelectric effect in $s+is$ phases can be
employed to obtain an advanced functionality in the practical
applications of certain iron-based superconductors. For this
purpose one can manipulate the symmetry of superconducting states
by simultaneously applying temperature bias and an external
magnetic field ${\bm H}_{ext}$ which in general removes the
degeneracy of $s+is/s-is$ states. As a result it is possible to
prepare a particular state by cooling the system through a BTRS
transition.
As an example we suggest a scheme of a memory cell based on the
non-trivial thermoelectric properties of bimetallic rings shown
in \Figref{Fig:Ring} with both branches made of $s+is$
superconductor, but having different doping. The junctions between
branches are maintained at different temperatures $T_{1,2}$ as
sketched in \Figref{Fig:Ring}(a).
%
%
%
%
%
There, the $s\pm is$ states have the opposite values of ${\bm B}_T$
\footnote{ The unconventional pattern of magnetic field
induced by counter-flows is related to the general property of
non-Meissner contribution to electrodynamics, in multicomponent
superconductors. The essence of the physics leading to the
Meissner and, in a broader sense, to the Anderson-Higgs effect is
that the magnetic field is described by a massive vector field
theory. As shown in Refs.~\cite{Babaev.Faddeev.ea:02,Babaev:09},
multicomponent systems cannot generically be described by a
massive vector field theory. Rather, it can be mapped onto a
massive vector field theory coupled to another field associated
with the charge-carrying counter-flows. The later terms lead to
contributions which are typically negligible or unobservable in
ordinary multiband superconductors due for example to interband
couplings. By contrast as discussed in this paper, the
thermoelectric effect in the $s+is$ state, makes the interband
phase fluctuations propagate into the bulk. This produces the
unconventional magnetic signatures.
}. \nocite{Babaev.Faddeev.ea:02,Babaev:09}
In such a system the value of the $\groupZ{2}$ index, which
distinguishes $s\pm is$ states, can be considered as a bit of
information. Let us now describe the writing/reading protocols.
First the external field ${\bm H_{ext}}$ interacts with thermally
induced currents and removes the $s\pm is$ degeneracy. Hence the
particular $s+is$ or $s-is$ state can be selected when such a system
is cooled through the BTRS phase transition, in the presence of both
external field and temperature bias $\delta T = T_2-T_1$. After cooling,
both the external field and temperature bias can be removed, leaving
the system in the ground state characterized by the $\groupZ{2}$ index
$\mathrm{sgn}(\delta T H_{ext})$. The read-out protocol, i.e. the
measurement of $\groupZ{2}$ index, can be implemented by applying
the temperature bias (in the absence of external field) and
measuring the direction of induced magnetic flux.

To conclude, we demonstrated that multicomponent superconductors
with broken time-reversal symmetry, and in particular the $s+is$
state, feature in the vicinity of BTRS transition, a giant
thermoelectric effect of principally different nature than that in
single-component superconductors. It originates in thermally
induced intercomponent counter-flows, in contrast to the
counter-flows of normal and superconducting currents in the
mechanism originally discussed by Ginzburg.
Although related effects should be present in various
multicomponent superconductors, along with multicomponent
superfluids, we focused on the $s+is$ states where the effect
is generic (irrespective of the model).

\begin{acknowledgments}
The work was supported by the Swedish Research Council grants 
642-2013-7837,  325-2009-7664.
The computations were performed on resources provided by the
Swedish National Infrastructure for Computing (SNIC) at National
Supercomputer Center at Link\"oping, Sweden.
\end{acknowledgments}

\appendix
\section{Ginzburg-Landau expansion for multiband superconductors}
\label{Appendix1}

To derive the Ginzburg-Landau expansion \Eqref{Eq:GL}, we use the
microscopic self-consistency equation
\begin{equation}\label{Eq:SelfConsistency1a}
{\bm \Delta } =2\pi T\hat\Lambda  \sum_{n=0}^{N_d} {\bm F (\omega_n) }\,,
\end{equation}
where  ${\bm F} = ( F_{1}, F_{2}, F_{3} )$  and $F_{k} (\omega_n)=
\Delta_k/\sqrt{\omega_n^2+ |\Delta_k|^2} $. $\omega_n=(2n+1)\pi
T$, where $n\in \mathbb{Z}$ is the fermionic Matsubara frequency,
$T$ the temperature, and $N_d=\Omega_d/(2\pi T)$ is a cut-off at
Debye frequency. In the diffusive case the anomalous functions
in each band can be expanded as follows:
\begin{equation}\label{Eq:GLexpansion}
F_j= \frac{\Delta_j}{\omega_n} -  \frac{\Delta_j |\Delta_j|^2}{2\omega_n^3}
    +\frac{ D_j (\Grad - ie\A)^2\Delta_j}{2\omega_n^2}\,,
\end{equation}
where $D_j$ is the diffusion coefficient. Then we get for the summation
over Matsubara frequencies
\begin{subequations}\label{Eq:Coefficients}
 \begin{eqnarray}
   2\pi T \sum_{n=0}^{\infty} 1/\omega_n &= G_0 + \tau      \,, \\
   2\pi T \sum_{n=0}^{\infty} 1/(2\omega_n^{3}) &= 0.1/T^2  \,, \\
   2\pi T \sum_{n=0}^{\infty} 1/(2\omega_n^{2}) &= 0.4 / T  \,,
 \end{eqnarray}
\end{subequations}
where $\tau = (1- T/T_c)$. We normalize $\Delta$ by $T_c$ and
$\xi_0= \sqrt{\pi D_0/8T_c}$, where $D_0$ is some arbitrary diffusion
constant.
\begin{equation}\label{Eq:GL3Band}
  (G_0 +\tau - \hat\Lambda^{-1} ) {\bm\Delta}
  - b |{\bm \Delta}|^2 \cdot {\bm \Delta}
  = {\bm{\tilde{D}}}\cdot ({\bm \Pi}^2 {\bm\Delta} )
\end{equation}
where $b=0.1$,  ${\bm \Pi} = \Grad - ie\A $,
${\bm \Delta} = (\Delta_1, \Delta_2, \Delta_3)$,
$|{\bm \Delta}|^2 = (|\Delta_1|^2, |\Delta_2|^2, |\Delta_3|^2) $,
and ${\bm{\tilde{D}}} = (\tilde{D}_1, \tilde{D}_2, \tilde{D}_3) $
where $\tilde{D}_j = D_j /D_0$. The diffusion coefficients can be
different. In the diffusive case the mixed gradient terms are absent.

Let us consider the case of the intraband dominated pairing which
can be described by a three-component GL theory in the vicinity of
$T_c$. This regime is described by the following coupling matrix:
\begin{equation}\label{Eq:Model3Band1}
\hat\Lambda =  \left(%
\begin{array}{ccc}
      \lambda & -\eta_h & -\eta_e \\
      -\eta_h & \lambda & -\eta_e \\
      -\eta_e & -\eta_e & \lambda
\end{array}%
\right)
\end{equation}
where $\eta_e, \eta_h \ll \lambda$. The critical temperature is
determined by the equation $G_0 = \min (\lambda^{-1}_1, \lambda^{-1}_2,
\lambda^{-1}_3)$, where $\lambda_{1,2}^{-1} = (2\lambda-\eta_h \pm
\sqrt{8\eta_e^2+\eta_h^2})/[2(\lambda^2 - \lambda\eta_h-2\eta_e^2)] $
and  $\lambda_{3}^{-1} = 1/(\lambda+\eta_h) $ are the positive eigenvalues
of the matrix
\begin{equation}\label{Eq:Model3Band2}
\hat\Lambda^{-1} =  X\left(%
\begin{array}{ccc}
    \lambda^2- \eta_e^2    & \eta_e^2+\lambda\eta_h  & \eta_e(\lambda+\eta_h) \\
    \eta_e^2+\lambda\eta_h & \lambda^2- \eta_e^2    & \eta_e(\lambda+\eta_h) \\
    \eta_e(\lambda+\eta_h) & \eta_e(\lambda+\eta_h)     & \lambda^2- \eta_h^2
\end{array}%
\right)
\end{equation}
where $X=1/[(\lambda^2- \lambda\eta_h- 2\eta_e^2)(\lambda+\eta_h)]$.
Since we assume that $\eta_{e,h} >0 $ and $\eta_{e,h} \ll \lambda $
the critical temperature is given  by the smallest eigenvalue
$G_0 = 1/(\lambda+\eta_h) $ so that
\begin{equation}
G_0 \hat I - \hat\Lambda^{-1} =  - \left(%
\begin{array}{ccc}
  a_1 & a_1 & a_2 \\
  a_1 &  a_1 & a_2 \\
  a_2 & a_2 &  a_3 \\
\end{array}%
\right)
\end{equation}
where
\begin{subequations} \label{Eq:GLcoeff}
\begin{eqnarray}
   a_1 &=& (\eta_e^2+\lambda\eta_h)/X \\
   a_2 &=& \eta_e(\lambda+\eta_h)/X\\
   a_3 &=& (2\eta_e^2-\eta_h^2+\lambda\eta_h)/X\,.
\end{eqnarray}
\end{subequations}
For example, for $\lambda=1$ and $\eta_e=0.1$, $\eta_h=0.2$ we get
$a_1= 0.2244$, $a_2= 0.1282$ and $a_3=0.1923$.  For $\lambda=1$
and $\eta_e=\eta_h=0.1$ we get $a_1=a_2= a_3 = 0.1136$.

Then we get the system of GL equations
\begin{eqnarray}\nonumber 
   (\tau - a_1)\Delta_1 -a_1 \Delta_2 -a_2\Delta_3 - b \Delta_1 |\Delta_1|^2 + \tilde{D}_1 {\bm \Pi}^2 \Delta_1 &=& 0
   \\ \nonumber
  (\tau - a_1) \Delta_2 -a_1 \Delta_1-a_2\Delta_3 - b \Delta_2 |\Delta_2|^2 + \tilde{D}_2 {\bm \Pi}^2 \Delta_2 &=& 0
   \\ \nonumber
   (\tau - a_3)\Delta_3 -a_2 \Delta_1- a_2\Delta_2 - b \Delta_3 |\Delta_3|^2 + \tilde{D}_3 {\bm \Pi}^2 \Delta_3 &=& 0
\end{eqnarray}
where $b=0.1$,  ${\bm \Pi} = \Grad - ie\A $ and
$\tilde{D}_j = D_j /D_0$. The diffusion coefficients can be
different. In the diffusive case the mixed gradient terms are
absent. The free energy functional whose variations gives the
Ginzburg-Landau equations
reads as
 \Align{Eq:EnergyExample}{
\F=\frac{\B^2}{2} &+\sum_{k=1}^3\Big\{
    \tilde{D}_k\left|\bm{\Pi}\Delta_k\right|^2
    +\alpha_k|\Delta_k|^2+ \frac{\beta_k}{2}|\Delta_k|^4  \Big\}
    \nonumber \\
    &+\sum_{k=1}^3\sum_{j=k+1}^3\eta_{kj}\Big\{
    \Delta_k^* \Delta_j+\Delta_j^* \Delta_k   \Big\}  \,,
 }
 where $\beta_k=b$, $\eta_{12}=a_1$, $\eta_{13}=\eta_{23}=a_2$, $\alpha_k = \alpha_k^0 (T/T_k -1)$,
 and $\alpha_k^0= 1-a_k$, $T_k= T_c (1-a_k)$.

\section{Details of simulations}
\label{Appendix2}
In order to investigate the response to the thermal gradients,
we minimize numerically the free energy \Eqref{Eq:EnergyExample},
in zero external field, with the standard condition that no current
flows through the sample's boundaries. In simulations we use the
microscopically derived coefficients \Eqref{Eq:GLcoeff}. The
theory is discretized within a finite element formulation
\cite{Hecht:12} and the minimization is performed using a non-linear
conjugate gradient algorithm.

\subsection*{Details for the bimetallic ring}

In \Figref{Fig:Ring} of the main text we consider the case of a
bimetallic that consists of two branches in the $s+is$ state,
but with different chemical composition. The difference in the
chemical composition is modelled by modulating the Ginzburg-Landau
coefficients \Eqref{Eq:GLcoeff} in the lower branch to be
85\% of their value in the upper branch:
$a_i^{\mbox{\tiny (lower)}}=0.85 a_i^{\mbox{\tiny (upper)}}$.
Thus the $a_i$ vary stepwise while passing from one branch to
the other.
The interfaces between both branches are maintained at different
temperatures $T_1$ and $T_2$. Here this is the whole interface
that is maintained at a given temperature, and thus the
temperature is a linear function of the polar angle.
This set-up determines the spatially inhomogeneous coefficients
$\alpha_k$ and $\eta_{kj}$ used in \Figref{Fig:Ring}.

%

\end{document}